\documentclass[a4paper,fleqn,usenatbib]{mnras}

\usepackage{newtxtext,newtxmath}

\usepackage[T1]{fontenc}
\usepackage{ae,aecompl}

\usepackage{graphicx}	
\usepackage{amsmath}	
\usepackage{amssymb}	

\usepackage{adjustbox} 
\usepackage{lipsum}

\newcommand{\Me}{$M_\oplus$}

\title[ALMA observations of TRAPPIST-1]{Searching for a dusty cometary belt around TRAPPIST-1 with ALMA}
\author[S. Marino et al.]{ S. Marino$^{1}$\thanks{E-mail:
    sebastian.marino.estay@gmail.com}, M. C. Wyatt$^{2}$, G. M. Kennedy$^{3}$, M. Kama$^{2}$, L. Matr{\`a}$^{4}$,
  \newauthor{A. H. M. J. Triaud$^{5}$ and Th. Henning$^{1}$}\\  
  $^{1}$Max Planck Institute for Astronomy, K\"onigstuhl 17, 69117 Heidelberg, Germany\\
  $^{2}$Institute of Astronomy, University of Cambridge, Madingley Road, Cambridge CB3 0HA, UK\\
  $^{3}$Department of Physics, University of Warwick, Gibbet Hill Road, Coventry, CV4 7AL, UK\\
  $^{4}$Harvard-Smithsonian Center for Astrophysics, 60 Garden Street, Cambridge, MA 02138, USA\\
  $^{5}$School of Physics \& Astronomy, University of Birmingham, Edgbaston, Birmingham, B152TT, UK\\
}

\date{Accepted XXX. Received YYY; in original form ZZZ}

\pubyear{2019}

\begin{document}
\label{firstpage}
\pagerange{\pageref{firstpage}--\pageref{lastpage}}
\maketitle

\begin{abstract}

  Low mass stars might offer today the best opportunities to detect
  and characterise planetary systems, especially those harbouring
  close-in low mass temperate planets. Among those stars, TRAPPIST-1
  is exceptional since it has seven Earth-sized planets, of which
  three could sustain liquid water on their surfaces. Here we present
  new and deep ALMA observations of TRAPPIST-1 to look for an
  exo-Kuiper belt which can provide clues about the formation and
  architecture of this system. Our observations at 0.88~mm did not
  detect dust emission, but can place an upper limit of 23$\mu$Jy if
  the belt is smaller than 4~au, and 0.15~mJy if resolved and 100~au
  in radius. These limits correspond to low dust masses of
  $\sim10^{-5}-10^{-2}$~\Me, which are expected after 8~Gyr of
  collisional evolution unless the system was born with a
  $>20$~\Me\ belt of 100~km-sized planetesimals beyond 40~au or
  suffered a dynamical instability. This $20$~\Me\ mass upper limit is
  comparable to the combined mass in TRAPPIST-1 planets, thus it is
  possible that most of the available solid mass in this system was
  used to form the known planets. A similar analysis of the ALMA data
  on Proxima~Cen leads us to conclude that a belt born with a mass
  $\gtrsim1$~\Me\ in 100~km-sized planetesimals could explain its
  putative outer belt at 30~au. We recommend that future
  characterisations of debris discs around low mass stars should focus
  on nearby and young systems if possible.

\end{abstract}

\begin{keywords}
    circumstellar matter - planetary systems - planets and satellites:
    dynamical evolution and stability - techniques: interferometric -
    methods: numerical - stars: individual: TRAPPIST-1.
\end{keywords}



\section{Introduction}
\label{sec:intro}


In recent years, the study of planetary systems around low mass stars
has received great attention. This is partly due to low mass planets
being easier to detect via transits around low mass stars, but also
because their occurrence rate is higher compared to planets around FGK
stars \citep[e.g.][]{Mulders2015, Hardegree2019}. Moreover, because of
the lower luminosity of M stars, close-in planets could harbour liquid
water in these systems. One example of such systems is TRAPPIST-1, a
M8 dwarf star at 12~pc hosting at least seven Earth-sized planets
\citep{Gillon2016, Gillon2017, Luger2017}, all within 0.06~au. Three
of these planets lie at a distance from the star where long-lived
liquid water could exist on their surfaces \citep{OMalley-James2017},
although constraints on the composition of these planets are still
very uncertain despite major efforts \citep[e.g.][]{deWit2016,
  deWit2018, Moran2018, Wakeford2019, Dorn2018, Grim2018,
  Burdanov2019}. The composition of these planets is highly dependent
on how these planets formed. For example, if they formed in situ these
planets might be water poor \citep[e.g.][]{Hansen2012}, while if they
formed further out and migrated in, as suggested by the near-resonant
chain, then these planets might contain significant amounts of water
\citep[e.g.][]{Cresswell2006, Terquem2007, Ormel2017,
  Schoonenberg2019}.



Moreover, volatile delivery through impacts of icy material formed
further out \citep[e.g.][]{Marino2018scat, Kral2018trappist1,
  Schwarz2018, Dencs2019} could also affect the composition of their
atmospheres and surfaces. How much icy material lies exterior to a
planetary system can be constrained by infrared observations which are
sensitive to circumstellar dust that is continually replenished
through the collisional break up of km-sized planetesimals
\citep[i.e. debris discs, see reviews by][]{Wyatt2008, Krivov2010,
  Hughes2018}. Thanks to Spitzer and Herschel, we know that at least
20\% of A--K type stars host exo-Kuiper belts that are orders of
magnitude brighter (and likely more massive) than the Kuiper belt
\citep[][]{Eiroa2013, Matthews2014pp6, Montesinos2016,
  Sibthorpe2018}. However, the constraints on planetesimal discs
around M type stars are poorer due to several factors, including small
grain removal processes \citep{Plavchan2005} and observational biases
and low sensitivity \citep[e.g.][]{Wyatt2008, Lestrade2009, Binks2017,
  Kennedy2018}.

Thanks to ALMA's unprecedented sensitivity at sub-mm wavelengths, it
is now possible to search for planetesimal discs around low mass stars
at greater depth, and constrain the architecture of planetary systems
around low mass stars. In this paper we report deep ALMA observations
of TRAPPIST-1. While previous to our observations there was no
evidence for the presence of dust around TRAPPIST-1, the efficient
planet formation in this system, its potentially young age (highly
unconstrained until recently) and its proximity made it an ideal
target to look for a planetesimal belt. This paper is organised as
follows. In \S\ref{sec:obs} we describe the ALMA observations and
place upper limits on dust emission levels. Then, in \S\ref{sec:dis}
we compare these dust upper limits with collisional evolution
models. Finally, in \S\ref{sec:con} we summarise our results and
conclusions.




\begin{figure}
  \centering
  \includegraphics[trim=0.2cm 0.2cm 0.2cm 0.2cm, clip=true, width=0.45\textwidth]{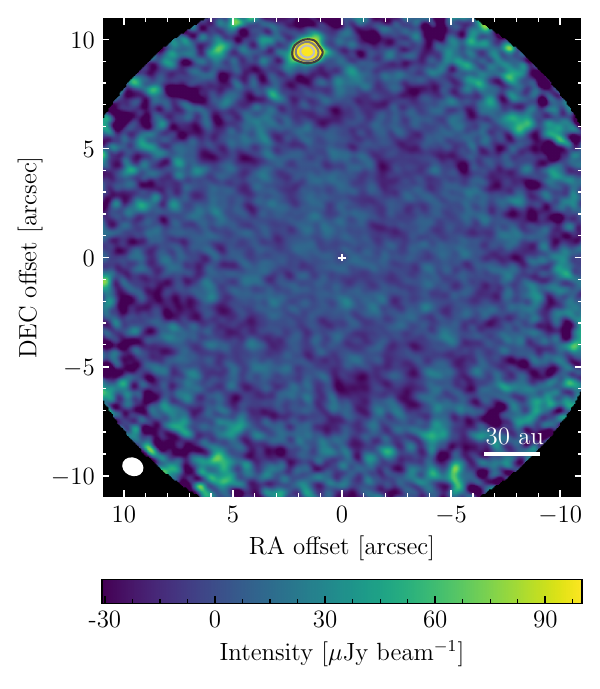}
 \caption{ALMA band 7 (0.88~mm) continuum image obtained with clean
   using natural weights. The beam size is $0\farcs71\times0\farcs54$
   and has a PA of $65\deg$. The image rms is
   7.7~$\mu$Jy~beam$^{-1}$. The beam size is represented by a white
   ellipse on the bottom left corner. The black regions at the edges
   of the image represent where the sensitivity drops below 20\% of
   that at the image center. The grey contours show emission above 5,
   10 and 20 times the rms level.}
 \label{fig:alma}
\end{figure}

\section{ALMA observations}
\label{sec:obs}

We observed TRAPPIST-1 using ALMA band 7 (0.88~mm) as part of the
project 2017.1.00215.S (PI: S. Marino). The observations were split
into 6 blocks that were executed between 3 May 2018 and 20 August
2018, with a total time on source of 4.5~h. Observations were taken
using a total of 44--48 antennas, with baselines ranging between 35
and 240~m (5$^\mathrm{th}$ and 80$^\mathrm{th}$ percentiles), which
allows to recover structure on angular scales ranging between
$0\farcs37$ and $5\arcsec$ (4.6 and 62 au projected in the sky). The
average PWV ranged between 0.4 and 0.7 between the 6 blocks. The
spectral setup was divided into 4 windows to observe the continuum,
centred at 334.6, 336.5, 348.5 and 346.6~GHz. The first three had a
total bandwidth of 2~GHz and a channel width of 15,625~kHz, while the
latter a total bandwidth of 1.875~GHz and a channel width of
488.281~kHz (effective spectral resolution of 0.845~km~s$^{-1}$) to
look for CO 3-2 line emission. Finally, the observations were
calibrated using CASA and the standard pipeline provided by ALMA.

We image the continuum with the task tclean in CASA, using natural
weights to produce a reconstructed image with the lowest possible
noise. Figure \ref{fig:alma} presents the clean image, corrected by
the primary beam and with a noise level or rms of
7.7~$\mu$Jy~beam$^{-1}$ at the center. We do not find any emission
arising from circumstellar material around TRAPPIST-1.  The only
detected source ($>5\sigma$) is a marginally resolved object with a
total flux 0.8~mJy that is likely sub-mm galaxy given ALMA number
counts \citep[we expect $\sim0.3$ sources within the primary beam with
  a flux equal or larger than 0.8~mJy,][]{Simpson2015, Carniani2015,
  Aravena2016}. We also search for extended emission from an edge-on
disk by computing the flux in a rectangular aperture centred on the
star of width $1.5\times$beam (since we assume an edge-on disc
co-planar with TRAPPIST-1 b--h) and variable length and position
angle; no $>3\sigma$ detection was found.




\begin{figure}
  \centering \includegraphics[trim=0.2cm 0.2cm 0.2cm 0.2cm, clip=true,
    width=1.0\columnwidth]{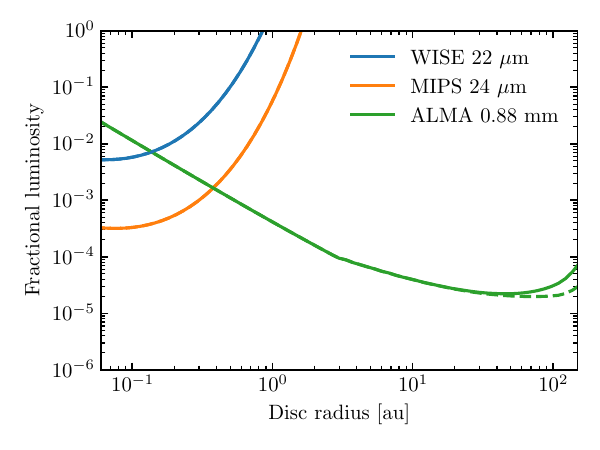}
 \caption{Upper limits on the fractional luminosity of a debris disc
   around TRAPPIST-1 based on WISE 22~$\mu$m (blue), MIPS 24~$\mu$m
   (orange) and ALMA 0.88~mm (green) data, assuming blackbody
   equilibrium temperatures (continuous lines). The dashed line
   represents the corrected limit by considering also the ISRF when
   calculating the equilibrium dust temperatures. Both green lines
   assume an edge-on disc orientation.}
 \label{fig:fvsr}
\end{figure}

We use the clean image to derive an upper limit on the 0.88~mm flux
from any hidden dust in the system below our detection threshold. For
an unresolved disc with a radius smaller than 4~au, we obtain a
$3\sigma$ limit of 23~$\mu$Jy. A planetesimal belt in the system could
be larger and resolved. For this case we derive a flux upper limit by
estimating the integrated flux uncertainty over a rectangular aperture
centred on the star as described above, also taking into account the
number of beams in this area and how the noise level increases away
from the phase center. This leads to a flux upper limit that increases
as a function of the disc diameter or aperture size, e.g. we find an
upper limit of 150~$\mu$Jy if it has an outer radius of 100~au. These
flux upper limits can be converted to an upper limit on the disc
fractional luminosity as a function of radius as shown in Figure
\ref{fig:fvsr}. For this, we assume the dust has blackbody equilibrium
temperatures $T_\mathrm{BB}(r)=42 (r/\textrm{1\ au})^{-1/2}$
\citep[$L_\star=5.2\times10^{-4}\ L_\odot$,][]{Filippazzo2015} and an
opacity that declines with wavelength as $1/\lambda$ beyond 200~$\mu$m
\citep[][]{Wyatt2008}. In the same figure we also overlay in blue the
upper limit derived from WISE 22~$\mu$m \citep[3~mJy,][]{Wright2010}
and MIPS 24~$\mu$m observations \citep[0.2~mJy, which are limited by
  calibration uncertainties of $\sim5\%$,][]{Gautier2007}. We find
that the ALMA limit is significantly more constraining than the WISE
and MIPS limit beyond 0.4~au. This is due to its high sensitivity and
the longer wavelength, being more sensitive to cold dust. Based on our
new observations we can rule out a debris disc with fractional
luminosity higher than $\sim2\times10^{-5}$ at a radius between
10--100~au.

Note that assuming blackbody equilibrium temperatures only based on
the stellar radiation might not be a good assumption since the
interstellar radiation field (ISRF) could contribute significantly to
the radiation field at tens of au. In order to take this into account,
we calculate the dust temperature as
$(T_\mathrm{BB}^4(r)+T_\mathrm{ISRF}^4)^{1/4}$, where
$T_\mathrm{ISRF}$ is a fixed equilibrium temperature due to the
ISRF. The equilibrium temperature of grains in the diffuse
interstellar medium has been studied extensively
\citep[e.g.][]{Li2001}, finding temperatures in the range 10--20~K for
dust grains smaller than 1~$\mu$m, significantly higher than the
blackbody equilibrium temperature of 3.6~K \citep[obtained by
  integrating the analytic expressions presented by][]{Mezger1982,
  Mathis1983}. Larger grains, however, have lower temperatures close
to blackbody since they have almost constant opacities at short
wavelengths that dominate the ISRF. We thus assume
$T_\mathrm{ISRF}=3.6$~K and incorporate this lower bound for dust
temperatures to all calculations in \S\ref{sec:dis}. The green dashed
line in Figure~\ref{fig:fvsr} is the corrected upper limit when taking
into account the ISRF.
 
\begin{figure}
  \centering \includegraphics[trim=0.0cm 0.0cm 0.0cm 0.0cm, clip=true,
    width=1.0\columnwidth]{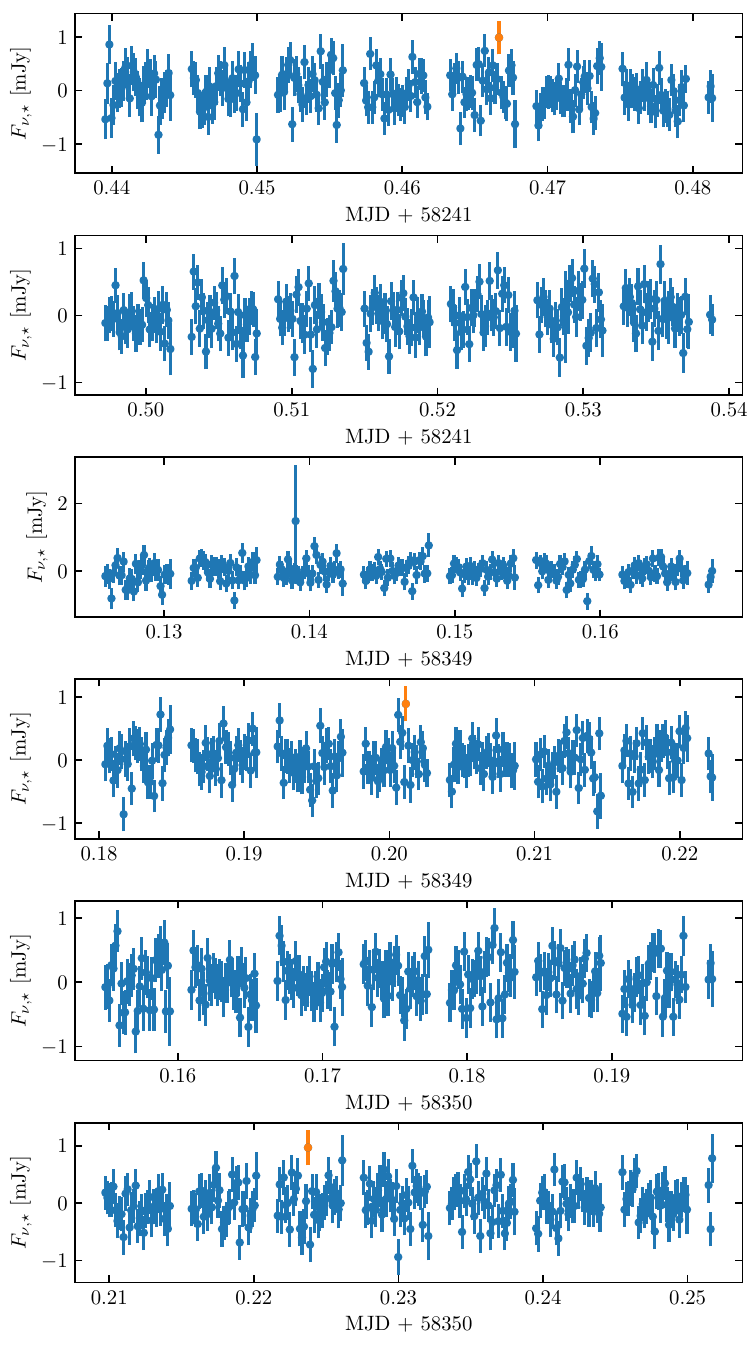}
 \caption{Measured flux at TRAPPIST-1 location vs time, integrated
   over 12s windows. Data points above $3\sigma$ are displayed in
   orange.}
 \label{fig:flares}
\end{figure}

We also look for any CO 3-2 line emission, which also led to a
non-detection. The achieved rms per 0.42~km~s$^{-1}$ channel is
0.5~mJy~beam$^{-1}$. Therefore we can set a $3\sigma$ upper limit of
6~mJy~km~s$^{-1}$ for any unresolved CO 3-2 emission interior to 4~au
and with a maximum Doppler shift of 10~km~s$^{-1}$ (i.e. beyond 1~au
if in Keplerian rotation). Using the non-LTE tool by \cite{Matra2018}
we translate this upper limit to a CO gas mass of
$\sim2\times10^{-8}$~\Me.

Because low mass stars can have strong and highly variable emission at
mm to cm wavelengths due to flaring activity, we also search for
unresolved variable emission at the stellar position by imaging the
data in time intervals of 12s. In Figure \ref{fig:flares} we show the
measured flux at the stellar position. We do not find any significant
variable emission above the noise level (rms of 0.23~mJy over 12s
window) that could arise from flares \citep[as in
  Proxima~Cen,][]{MacGregor2018prox}. We only find three single
$3\sigma$ peaks over 1400 data points which is roughly consistent with
the expected number of false positives for a normal distribution. The
non-detection of variable emission from TRAPPIST-1 is consistent with
results reported by \cite{Hughes2019} which did not detect any
emission at 3~mm with ALMA nor at 7~mm with the VLA for
TRAPPIST-1. Note that the limits presented here at 0.88~mm are still
consistent with flaring levels similar to Proxima~Cen when taking into
account the integration length and larger distance to TRAPPIST-1.

\section{Discussion}
\label{sec:dis}

In this section we aim to constrain what initial planetesimal belt
properties are still consistent with the observational limits and what
we can rule-out. We do this by comparing a collisional evolution model
to our non-detection of a debris disc around TRAPPIST-1
(\S\ref{sec:coll_trap}) and to archival ALMA observations of
Proxima~Cen (\S\ref{sec:coll_prox}).

We use the same collisional evolution model that has been used to fit
observations of resolved debris discs \citep{Wyatt2011,
  Marino201761vir}. This model solves the evolution of the size
distribution of solids in a collisional cascade, and here we assume
the following:
\begin{enumerate}
  \item a maximum planetesimal size of 100~km,
  \item size dependent strengths \citep{Benz1999, Stewart2009},
  \item internal densities of 2.7~g~cm$^{-3}$,
  \item mean orbital eccentricities of 0.05 and inclinations of
    $1.4\deg$ ($i\sim e/2$) that set the relative velocities,
  \item an initial surface density of solids equivalent to a Minimum
    Mass Solar Nebula (MMSN) extrapolated out to 100~au
    \citep[$\Sigma(r)=(r/1~\mathrm{au})^{-1.5}\ M_\oplus$~au$^{-2}$,][]{Weidenschilling1977mmsn,
      Hayashi1981}.
\end{enumerate}

For more detail on assumptions of this model we direct the reader to
\cite{Wyatt2011} and \cite{Marino201761vir}. The output from these
simulations are size distributions at different radii that we
translate to surface densities or disc masses in mm-sized grains, here
defined as all grains smaller than 1~cm. These mm-sized grains are
what our ALMA observations are most sensitive to, since grains in the
range 0.1--10~mm dominate the emission at these wavelengths and the
mass of grains smaller than 1~cm. This choice is consistent with the
dust opacity that is assumed to translate fluxes to dust masses (see
below).

\subsection{TRAPPIST-1 collisional evolution}
\label{sec:coll_trap}

Given the estimated age of 7.6$\pm$2.2~Gyr \citep{Burgasser2017}, it
is likely that any debris disc present around TRAPPIST-1 has suffered
significant collisional evolution. This means that even if this system
was born with a massive disc of planetesimals and detectable dust
levels, after 8~Gyr of evolution it could have lost most of its mass
through collisions and the removal of small dust subject to stellar
winds and radiation pressure (although the latter is not high enough
to remove grains larger than $\sim0.01~\mu$m).

To quantify which initial planetesimal belt parameters are allowed by
our non-detection, we derive an upper limit on the surface density and
total mass of a collisionally produced dust disk by assuming:
\begin{enumerate}
\item the belt is edge-on, i.e. co-planar to TRAPPIST-1 b--h,
\item a dust opacity of 3~cm$^{2}$~g$^{-1}$ at 0.88~mm,
\item equilibrium blackbody temperatures in the optically thin regime
  (considering both the stellar radiation and ISRF),
\item a belt width that is half of the belt central radius
  \citep[typical of debris discs,][]{Matra2018mmlaw}.
\end{enumerate}
The adopted dust opacity is consistent with the one expected for
grains smaller than 1~cm in a size distribution according to our
collisional evolution model, i.e. $N(a)\sim a^{-3.4}$ in the range
10~$\mu$m---100~m \citep{Woitke2016}.

\begin{figure}
  \centering \includegraphics[trim=0.2cm 0.2cm 0.2cm 0.2cm, clip=true,
    width=1.0\columnwidth]{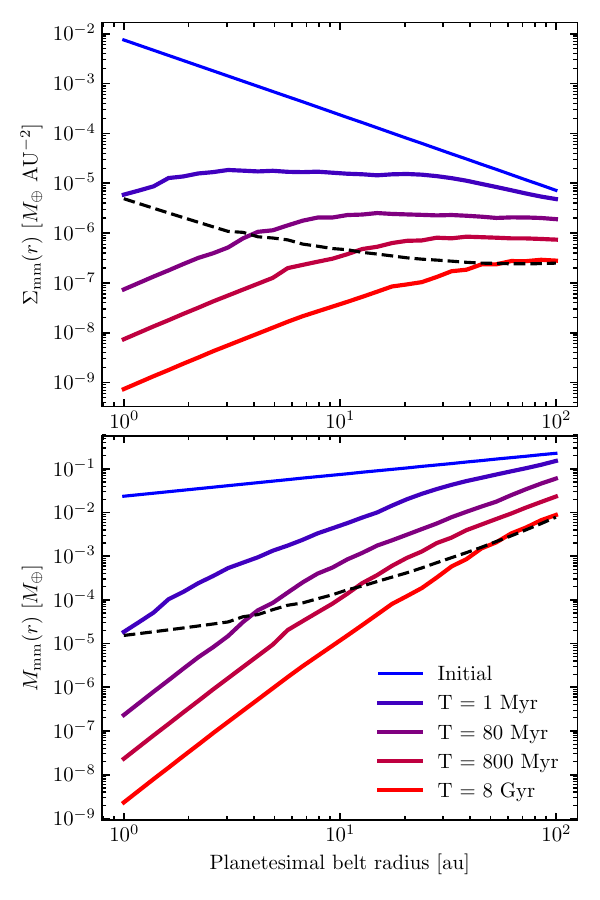}
 \caption{Collisional evolution of the surface density (top) and belt
   mass (bottom) in mm-sized grains around TRAPPIST-1, as a function
   of radius and time (continuous colour lines). As a comparison, the
   dashed line shows the ALMA upper limit derived in this work. The
   upper limit for the surface density is derived assuming a belt
   fractional width of 0.5 and an edge-on orientation (i.e., the
   dashed line shows the limit on the surface density of a belt at
   that radius, rather than the limit on the surface density of an
   extended disk at that radius).}
 \label{fig:coll}
\end{figure}

Figure \ref{fig:coll} compares the predicted surface density (top) and
disc mass in mm-sized grains at different epochs (continuous lines)
with the $3\sigma$ upper limit derived by our observations (dashed
lines). Note that our upper limits for the surface density do not
correspond to expected disc profiles, but rather the maximum surface
density if the planetesimal belt was centred at that radius and had a
fractional width (width over central radius) of 0.5. Beyond 40~au, our
model predicts dust levels that would be detectable. Therefore we
conclude that our observations rule-out that TRAPPIST-1 was born with
a planetesimal belt of mass similar to or larger than a MMSN
($\gtrsim20$~\Me) at a radius between 40--100~au. If we take the mass
and orbits of TRAPPIST-1 planets derived by \cite{Grim2018} we find a
\textit{minimum mass TRAPPIST-1 nebula} that is $\Sigma=670\pm40
(r/0.02~\mathrm{au})^{-1.8\pm0.1}\ M_\oplus$~au$^{-2}$. This
expression extrapolated to large radii translates into surface
densities lower than assumed in our model and thus consistent with our
non-detection. Note that this approach to derive an initial mass in
solids assumes planets' orbits do not evolve significantly, which
might not be the case for TRAPPIST-1 planets \citep{Ormel2017,
  Schoonenberg2019}. Moreover, extrapolating the relationship of
stellar luminosity and belt radius found by \cite{Matra2018} to
TRAPPIST-1's luminosity, we do not expect a belt radius larger than
25~au. Interior to 40~au, our upper limit is significantly higher than
the predicted dust levels after 8~Gyr of evolution, hence collisional
evolution alone can explain our disc non-detection, and thus a
planetesimal disc more massive than a MMSN could have formed there
when this system formed. Note that interior to 40~au this system could
still host a MMSN-like planetesimal disc but that is faint today due
to how the size distribution has evolved, with the mass in small
bodies orders of magnitude more depleted than the mass in the largest
bodies \citep{Schuppler2016, Marino201761vir}.

These conclusions hold when varying model parameters such as the
maximum planetesimal size and level of stirring since the surface
density of mm-sized grains in collisional equilibrium is not very
sensitive to these \citep[see Equation 3 in][]{Marino2019}. Other
assumptions to translate flux upper limits to surface densities could
have a small effect. For example, assuming the belt is wider would
distribute roughly the same dust mass upper limit (or flux upper
limit) over a larger area, and thus it would lower our upper limit on
the surface density derived from observations. Only by decreasing the
initial surface density of solids in our collisional model or
narrowing the assumed belt, the predicted surface density of mm-sized
dust would be lower than our upper limit at all radii. A caveat in the
use of our collisional evolution model is that it assumes the system
has been stable for 8~Gyr. The non-detection of a disc could also be
explained by an instability in the system that scattered and depleted
a massive planetesimal belt.


\begin{figure}
  \centering \includegraphics[trim=0.2cm 0.2cm 0.2cm 0.2cm, clip=true,
    width=1.\columnwidth]{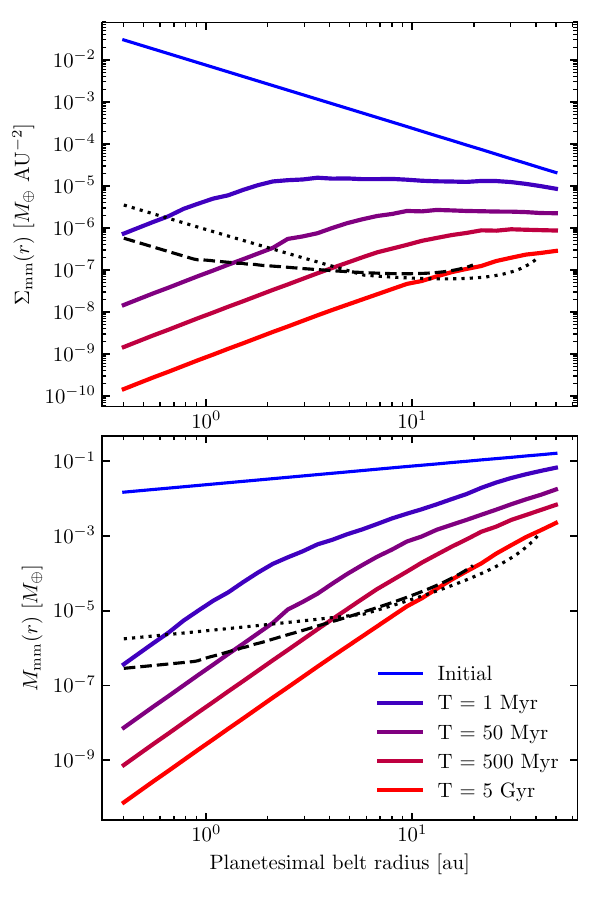}
 \caption{Collisional evolution of the surface density (top) and belt
   mass (bottom) in mm-sized grains around Proxima~Cen, as a function
   of radius and time (continuous colour lines). As a comparison, the
   dashed and dotted lines show the ALMA upper limit derived using the
   12m+ACA and ACA alone images, respectively. The upper limit for the
   surface density is derived assuming a belt fractional width of 0.5
   and a belt inclination of $45\deg$ as suggested by
   \citet{Anglada2017}.}
 \label{fig:proxima}
\end{figure}


\subsection{Proxima~Cen collisional evolution}
\label{sec:coll_prox}

We applied the same model to Proxima~Cen \citep[age of
  $\sim5$~Gyr,][]{Bazot2016} which hosts a low mass temperate planet
\citep{Anglada2016} and has also been observed by ALMA at 1.3~mm. To
derive upper limits we re-imaged the data after applying the
calibration script provided by ALMA. In the original analysis of the
data by \cite{Anglada2017}, they proposed the existence of warm dust
component at 0.4~au, a cold belt at 1--4~au and an outer belt at
30~au. An independent analysis by \cite{MacGregor2018prox} of the same
observations showed that in the same ALMA data there is strong and
time variable flaring activity from Proxima~Cen. This time-variable
and unresolved emission could have misled \cite{Anglada2017} to
conclude that there is circumstellar dust within a few
au. Moreover, the signal from the putative outer belt after
  azimuthally averaging \citep[as in][]{Marino2016} is only marginally
  significant at $3.5\sigma$. A future detailed analysis of more
  sensitive observations in the visibility space, taking into account
  the time variable emission of Proxima~Cen, should be able to confirm
  or rule out the presence of dust emission in this system at the
  levels claimed by \cite{Anglada2017}.


Despite this ongoing debate on the presence of debris-like dust around
Proxima~Cen, in Figure \ref{fig:proxima} we compare the results from
collisional evolution with the upper limits from the ACA map (dotted
line) and 12m and ACA combined (dashed line) assuming a disc inclined
by $45\deg$ and azimuthally averaging. We find that the ACA
observations could have marginally detected a disc if it had a mean
radius between 10--40~au and was as massive as a MMSN (10--20~\Me)
under the assumptions stated above. This mass however is not well
constrained since interior to 40~au the whole size distribution is in
collisional equilibrium \citep[proved by the constant slope of
  $r^{7/3}$ in the model surface density,][]{Wyatt2008,
  Kennedy2010}. This means that even if the belt had started with a
larger mass it would have depleted faster reaching the same mass after
5~Gyr. By varying the initial disc mass we find that discs with a mass
lower than a tenth of a MMSN would have a surface density below the
detection limit (at $\sim30$~au). Therefore, we conclude that the
amount of mm-sized dust in the putative outer belt around Proxima~Cen
is not unrealistic, and roughly consistent with the collisional
evolution of a planetesimal belt that was born with a tenth of a MMSN
(1--2~\Me).

\subsection{Searching for debris-like dust around low mass stars}

\label{sec:sensitivity}
In this paper we have shown how important collisional evolution is
when interpreting upper limits on the presence of dust around low mass
stars. At tens of au, the dust mass is expected to decrease with age
roughly as $t^{-0.4}$ while the lifetime of the largest planetesimals
is longer than the age of the system, and $t^{-1}$ at later times
\citep[e.g.][]{Lohne2008}. Thus, the age of surveyed systems is a key
factor to consider when selecting targets to observe. In addition to
this, the distance to the source is very important too since the flux
is inversely proportional to the distance squared. Proof of this is
that the sensitivities or upper limits on the dust mass around
Proxima~Cen are $\sim5$ times better than for TRAPPIST-1, even though
the noise in the reconstructed images was poorer. Future searches
should take into account both age and distance to be the most
sensitive to debris-like dust. Therefore, we recommend then that
surveyed samples should be composed of young and nearby systems that
are expected to have the largest flux, i.e. that minimise the quantity
$t^{0.4} d^{2}$ if the largest planetesimals are not yet in
collisional equilibrium \citep[$r>r_\textrm{c}$ in Equation 8
  from][]{Marino201761vir}, or $t d^{2}$ if they are
($r<r_\textrm{c}$).


\section{Summary and conclusions}
\label{sec:con}

In this paper we have reported new ALMA observations around TRAPPIST-1
at 0.88~mm, the deepest to search for dust emission from a debris disc
around this system. These observations did not detect circumstellar
dust or CO 3-2 emission. We compared our dust upper limits with
collisional evolution models, which given TRAPPIST-1's age of
$\sim8$~Gyr predict detectable dust levels beyond 40~au if the initial
disc was as massive as a MMSN. Therefore our model rules-out that
TRAPPIST-1 was born with a planetesimal belt larger than 40~au and
with a mass similar or higher than a MMSN ($\gtrsim20$~\Me). Within
40~au, on the other hand, the surface density or total mass of
mm-sized dust could be simply depleted due to collisional evolution
and avoid detection. The solid mass upper limit derived here is
comparable to the mass in TRAPPIST-1 planets
\citep[$\sim5.7$~\Me,][]{Grim2018}, thus it possible that most of the
available solid mass in these systems was transported inwards and used
to form the known planets.

We searched in time bins of 12s for any flaring activity of TRAPPIST-1
that could be present in this data. We did not find any significant
emission, and we derive a $5\sigma$ upper limit of 1.2~mJy for
variable emission over 12s windows.

Given the available archival ALMA data on Proxima~Cen, also a system
around a low mass star hosting a low mass temperate planet, we
performed a similar analysis. We compared our model with ALMA
observations and showed that the current upper limits are slightly
below the mass of mm-sized dust that we expect given our collisional
evolution model. The archival observations could have marginally
detected a belt with a mean radius between 10--40~au and with an
initial mass $\gtrsim1$~\Me. This means that the marginal detection of
an outer belt at 30~au by \cite{Anglada2017} is consistent with our
collisional evolution model and a belt that was born with a tenth of a
MMSN. An even more massive disc would have collisionally evolved to
the same mass and thus these observations cannot constrain well the
initial mass of a putative planetesimal belt. Interior to 10~au, the
limits cannot rule out that the system was born with a planetesimal
belt more massive than a MMSN.

We conclude that in order to set tighter constraints on planetesimal
discs around low mass stars with planets, we should focus most efforts
on nearby and, if possible, young ($\lesssim1$~Gyr old)
systems. Nearby systems ensure a higher flux with a strong dependence
on the distance, while younger systems are also more likely to host
not yet collisionally depleted belts beyond a few au. In
\S\ref{sec:sensitivity} we showed how the flux depends more strongly
on the distance, and how both distance and age can be taken into
account when prioritising which targets to observe. Such systems could
provide important constraints to the formation of planetary systems
around low mass stars.
  
\section*{Acknowledgements}

GMK is supported by the Royal Society as a Royal Society University
Research Fellow. LM acknowledges support from the Smithsonian
Institution as a Submillimeter Array (SMA) Fellow. MK gratefully
acknowledges funding from the European Union's Horizon 2020 research
and innovation programme under the Marie Sklodowska-Curie Fellowship
grant agreement No 753799. T.H. acknowledges support from the European
Research Council under the Horizon 2020 Framework Program via the ERC
Advanced Grant Origins 83 24 28. This paper makes use of the following
ALMA data: ADS/JAO.ALMA\#2017.1.00215.S and
ADS/JAO.ALMA\#2016.A.00013.S. ALMA is a partnership of ESO
(representing its member states), NSF (USA) and NINS (Japan), together
with NRC (Canada), MOST and ASIAA (Taiwan), and KASI (Republic of
Korea), in cooperation with the Republic of Chile. The Joint ALMA
Observatory is operated by ESO, AUI/NRAO and NAOJ.




\bibliographystyle{mnras} \bibliography{SM_pformation} 






\bsp	
\label{lastpage}
\end{document}